\documentclass[twocolumn,twoside,slac_two]{revtex4}
\usepackage{graphicx}
\usepackage{fancyhdr}
\begin{document}

%Title of paper
\title{$W$ and $Z$ Boson Production at Hadron Colliders}

\author{C. Hays}
\affiliation{Department of Physics, University of Oxford, Oxford OX1 3RH, United Kingdom}

\begin{abstract}
The electroweak theory has been tested to high precision, with measurements probing
its predictions at the loop level.  The current generation of particle accelerators will 
produce enough $W$ and $Z$ bosons through hadron collisions to significantly improve the 
accuracy of these measurements.  I review the issues related to such production, with 
particular emphasis on associated uncertainties on the $W$ boson mass, which has now been 
measured more precisely at the Tevatron than at the Large Electron Positron collider.
\end{abstract}

\maketitle

\thispagestyle{fancy}

%%%%%%%%%%%%%%%%%%%%%%%%%%%%%%%%%%
\section{Introduction}

The electroweak theory is highly overconstrained, with three fundamental parameters at 
tree level \cite{ewk} and more than a dozen precise measurements of quantities derived 
from these parameters \cite{ewkmeasurements}.  The experimental precision of these 
measurements, typically at the $0.1\%$ level, is sufficient to probe for loop interactions 
of both observed and unobserved particles.  

Ongoing and future measurements at the Fermilab Tevatron and the Large Hadron Collider 
(LHC) will improve the accuracy of several quantities at the loop level.  The mixing 
angle between the electromagnetic and weak symmetries, accessed through forward-backward 
lepton asymmetries, could have a reduced uncertainty from the full Tevatron data set 
\cite{d0afb}.  The $W$ boson mass and width have been measured most precisely by the 
D\O\ \cite{d0wmasswidth} and CDF \cite{cdfwmasswidth} experiments, respectively, with 
accuracies that will significantly improve with the larger available data sets.  The 
combined Tevatron top-quark mass measurement has a relative precision of 0.75\% 
\cite{topmass}.

Of these measurements, the $W$ boson mass has the greatest potential in the near term 
to significantly tighten constraints on unobserved particles \cite{cdfwmasswidth}.  The 
combination of existing measurements gives $m_W = 80.399 \pm 0.023$ GeV \cite{combination}, 
and future CDF and D\O\ measurements using data already collected will be at least this 
precise.  Including predictions from the LHC, hadron-collider measurements expect $<10$ MeV 
precision \cite{atlas}, or about a factor of three reduction of the current uncertainty.

A reduction of $m_W$ uncertainty will directly constrain the properties of new particles. 
The tree-level prediction $m_W = 79.964 \pm 0.005$ GeV is more than $18 \sigma$ from the 
measured value.  An important loop correction arises from the top-bottom loop, due to 
the large mass difference $m_t - m_b$, with the correction proportional to $m_t^2$ 
\cite{corrections}.  The correction arising from Higgs boson loops is proportional to 
$\ln m_H$.  Table~\ref{tbl:inputs} shows the shift in $m_W$ due to a doubling of $m_H$ 
and to $+1\sigma$ shifts in several inputs to $m_W$ \cite{cdfwmasswidth,dmdpar}.

\begin{table}[ht]
\begin{center}
\caption{The $m_W$ shift due to a factor of two change in $m_H$, or $+1\sigma$ shifts 
in the input parameters $m_t$, $m_Z$, and $\alpha_{EM}$ \cite{cdfwmasswidth,dmdpar}. }
\begin{tabular}{|l|c|}
\hline
Parameter Shift                                                & $m_W$ Shift \\
                                                               & (MeV/$c^2$) \\
\hline
$\Delta \ln m_H = + 0.693$                                     &  -41.3 \\
$\Delta m_t = + 1.3$ GeV/$c^2$                                 & 7.9   \\
$\Delta \alpha_{EM}(Q = m_Z c^2) = + 0.00035$                  & -6.2   \\
$\Delta m_Z = + 2.1$ MeV/$c^2$                                 & 2.6    \\
\hline
\end{tabular}
\label{tbl:inputs}
\end{center}
\end{table}

Given the ongoing and potential constraints from measurements of $m_W$, I focus 
on the status of experimental and theoretical uncertainties on this measurement at 
hadron colliders.

%%%%%%%%%%%%%%%%%%%%%%%%%%%%%%%%%%
\section{$W$ and $Z$ Boson Production}

\begin{figure}[h]
\centering
\includegraphics[width=85mm]{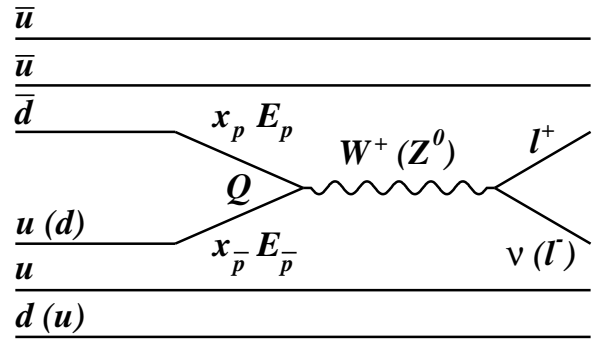}
\caption{Leading-order production of $W$ and $Z$ bosons at the Tevatron \cite{cdfwmasswidth}.  
Additional corrections from initial-state QCD or final-state QED radiation must be modelled 
accurately for the $m_W$ measurement. } \label{fig:wzprod}
\end{figure}

There are many components of $W$ and $Z$ boson production at the Tevatron that enter 
into the $m_W$ measurement (Fig.~\ref{fig:wzprod} \cite{cdfwmasswidth}).  The interacting 
partons have a fraction $x$ of the (anti)proton's momenta, with the relative fractions 
determining the boson's longitudinal momentum.  Initial-state radiation (ISR) of gluons 
or photons can give the boson a transverse boost $p_T^{W (Z)}$.  The boson decay is governed 
by the lepton electroweak charges, and spin correlations with the QCD ISR affect the decay 
angles.  Finally, final-state radiation reduces the momentum of the charged lepton(s), 
affecting the inferred boson mass.

%%%%%%%%%%%%%%%%%%%%%%%%%%%%%%%%%%
\subsection{Parton Distribution Functions}

The momentum fraction $x$ of a given colliding parton is described by the parton 
distribution functions (PDFs).  The PDFs are defined and fit to global data by 
independent groups \cite{cteq,mrst} at a fixed momentum transfer $Q$, and extrapolated 
to higher $Q$ using the DGLAP equations \cite{dglap}.  Uncertainties on the input data 
are typically smaller than the deviations between the data for a given parton distribution 
function, resulting in poor global values of $\chi^2$ on the fits.  Estimates of the PDF 
uncertainty on any given quantity follow an ad-hoc recipe defined by the fitters.  The 
recipe typically gives a ``90\% confidence level (C.L.)'' uncertainty, though this is 
based more on experience than on pure statistics.  A significant challenge to the $m_W$ 
measurement is ensuring the accuracy of this uncertainty, and reducing it.  

The uncertainty on $m_W$ due to PDFs arises from the detector acceptance to a 
charged lepton at a given pseudorapidity.  Charged leptons decaying transverse 
to the beam carry the highest $p_T$ (half the boson mass, to first order).  The 
smaller the decay angle with respect to the beam line, the smaller the $p_T$.  
At small angles, the charged lepton leaves the detector acceptance, and boosting 
the lepton along the beam line affects the distribution of angles (and, 
correspondingly $p_T$) accepted by the detector.  An uncertainty on the boost
translates weakly into an uncertainty on $m_W$.  The PDF uncertainty on the 
most recent D\O\ (CDF) $m_W$ measurement is 10 (13) MeV \cite{d0wmasswidth,cdfwmasswidth}.

Tevatron data provide significant constraints on PDFs.  For the $m_W$ measurement, 
the most relevant constraints come from measurements of the $Z$ boson rapidity 
distribution and the $W$ boson production charge asymmetry.  Because 
$\sigma_W \times BR(W\rightarrow l\nu)/ \sigma_Z \times BR(Z\rightarrow ll) \approx 10$ \cite{cdfxsec}, 
the $W$ boson charge asymmetry has greater statistical power than the $Z$ boson 
rapidity.  In addition, the charge asymmetry is a direct study of the $W$ boson 
production relevant to the $m_W$ measurement.

%%%%%%%%%%%%%%%%%%%%%%%%%%%%%%%%%
\subsubsection{$W$ Boson Charge Asymmetry}

\begin{figure}[h]
\centering
\includegraphics[width=75mm]{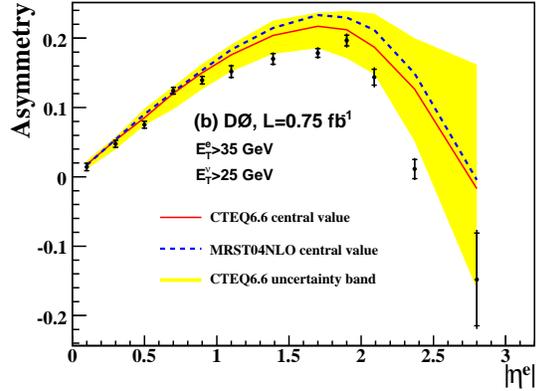}
\caption{The electron charge asymmetry as a function of pseudorapidity, as measured 
by the D\O\ Collaboration (points) and predicted by the CTEQ (solid line) and MRST 
(dashed line) PDF fits.  Also shown is the CTEQ 90\% C.L. uncertainty band. }
\label{fig:D0WAsymmetry}
\end{figure}

On average, up quarks carry a higher fraction of the proton's momentum than down 
quarks.  Thus, the longitudinal momentum of the $W^+$ boson tends to be in the 
direction of the proton momentum.  A measurement of the asymmetry between $W^+$ 
and $W^-$ production at a given boson rapidity gives information on the ratio 
of up- to down-quark momentum fraction.  Historically, the asymmetry between 
the charged leptons from the $W$ boson decay have been measured, since the neutrino 
is not measured and the $W$-boson's rapidity can not be fully reconstructed.  The 
D\O\ Collaboration has recently performed such a measurement 
(Fig.~\ref{fig:D0WAsymmetry} \cite{d0wasym}), and incorporating its results into 
the PDF fits will improve their accuracy.

\begin{figure}[h]
\centering
\includegraphics[width=75mm]{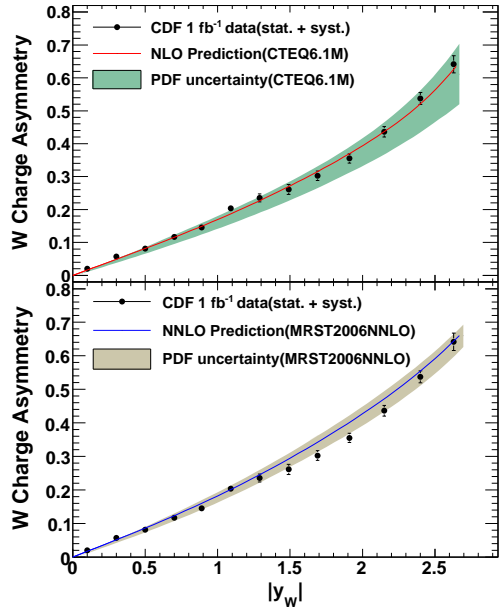}
\caption{The $W$ boson charge asymmetry as a function of rapidity, as measured by 
the CDF Collaboration (points) and predicted by the CTEQ (top) and MRST (bottom) 
fits (including the 90\% C.L. uncertainty bands). } 
\label{fig:CDFWAsymmetry}
\end{figure}

The CDF Collaboration has developed a novel method for directly measuring the 
$W$-boson charge asymmetry.  The method solves for the boson rapidity using the 
$W$-boson mass as a constraint.  The two solutions are given weights according to 
the expected boson kinematic and decay distributions.  To remove any dependence on 
the input charge asymmetry, the procedure is iterated until a stable solution is 
reached.  The CDF data (Fig.~\ref{fig:CDFWAsymmetry} \cite{cdfwasym}) will 
significantly improve predictions of the PDFs at high $W$ boson rapidity.
%
%The new CDF and D\O\ results have yet to be incorporated into the PDFs and the 
%$m_W$ measurements.  With the new results the corresponding uncertainties on the 
%$m_W$ measurements should be reduced.
%

%%%%%%%%%%%%%%%%%%%%%%%%%%%%%%%%%%
\subsubsection{Issues for the $m_W$ Measurement}

With the overall precision on $m_W$ expected to approach 20 MeV in the next 
iteration of Tevatron measurements, it is useful to consider methods to produce 
a more robust estimate of the PDF uncertainty.  Currently, both CDF and D\O\ 
rescale the 90\% C.L. uncertainty obtained from the CTEQ recipe to produce a 
68\% C.L. uncertainty on $m_W$.  This is motivated by the empirical observation 
that the spread of data for the valence $u$ and $d$ quarks are roughly Gaussian 
\cite{pdfscaling}.  However, the gluon contribution to the $m_W$ uncertainty is 
non-negligible, and the 90\% C.L. definition is not obtained strictly by statistics.  
There is thus some ambiguity as to whether the rescaling of uncertainties is 
appropriate.

In addition to the question of uncertainty scaling, there is the issue of the 
functions used to parametrize the PDFs.  There is considerable flexibility in 
the choice of functions, and there are various assumptions that are generally 
made to reduce the number of parameters in the fits.  The existence of multiple 
PDF fits is extremely useful in this regard, and with respect to the definition 
of the uncertainty, since the various fits use different parametrizations.  
However, the existing fits are clearly not exhaustive, raising the possibility 
of an underestimated uncertainty due to a parametrization that poorly describes 
the distribution function.

There are several possible strategies for obtaining a more robust PDF uncertainty. 
One can measure $m_W$ using leptons at higher pseudorapidity, though this involves 
enormous effort to calibrate these detector regions.  One can fit for $m_W$ in 
several lepton pseudorapidity bins to demonstrate that the PDFs accurately describe 
the distributions within the detector acceptance.  Or one can apply an uncertainty 
obtained strictly (or dominantly) from Tevatron data, which would presumably provide 
a better $\chi^2$ but a larger uncertainty.  At the LHC, the larger statistics and 
detector coverage will make some of these tests more feasible than at the Tevatron.

%%%%%%%%%%%%%%%%%%%%%%%%%%%%%%%%%%
\subsection{Boson $p_T$}

\begin{figure}[h]
\centering
\includegraphics[width=90mm]{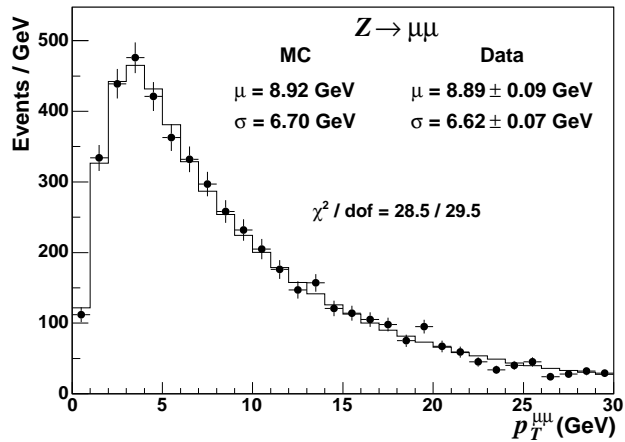}
\caption{The measured dimuon $p_T$ in 200 pb$^{-1}$ of CDF data, for muon pairs with 
invariant mass between 66 and 116 GeV.}
\label{fig:cdfzpt}
\end{figure}

A majority of $W$ and $Z$ bosons are produced with low $p_T$ (Fig.~\ref{fig:cdfzpt} 
\cite{cdfwmasswidth}), where non-perturbative QCD must be used to describe the $p_T$ 
distribution.  Both CDF and D\O\ model this distribution using the {\sc resbos} 
generator \cite{resbos}, which is based on a differential calculation with parameters 
motivated by a resummation calculation in the non-perturbative regime.  There are three 
parameters, $g_i$ ($i=1,2,3$), whose values are constrained by fits to data.  The most 
relevant parameter for the $m_W$ measurement is $g_2$, which determines the position of 
the distribution's peak.  

CDF obtains $g_2 = 0.685 \pm 0.048 (\rm stat)$ \cite{cdfwmasswidth} using the 
$Z\rightarrow ll$ control samples for its $m_W$ measurement.  This value of $g_2$, 
which uses CTEQ6M PDFs, is consistent with the value of $g_2 = 0.68^{+0.02}_{-0.01}$ 
obtained from a global fit using CTEQ3M PDFs.  The other $g_i$ parameters are correlated 
and CDF found that varying $g_3$ has a negligible effect on $m_W$.  

\begin{figure}[h]
\centering
\includegraphics[width=85mm]{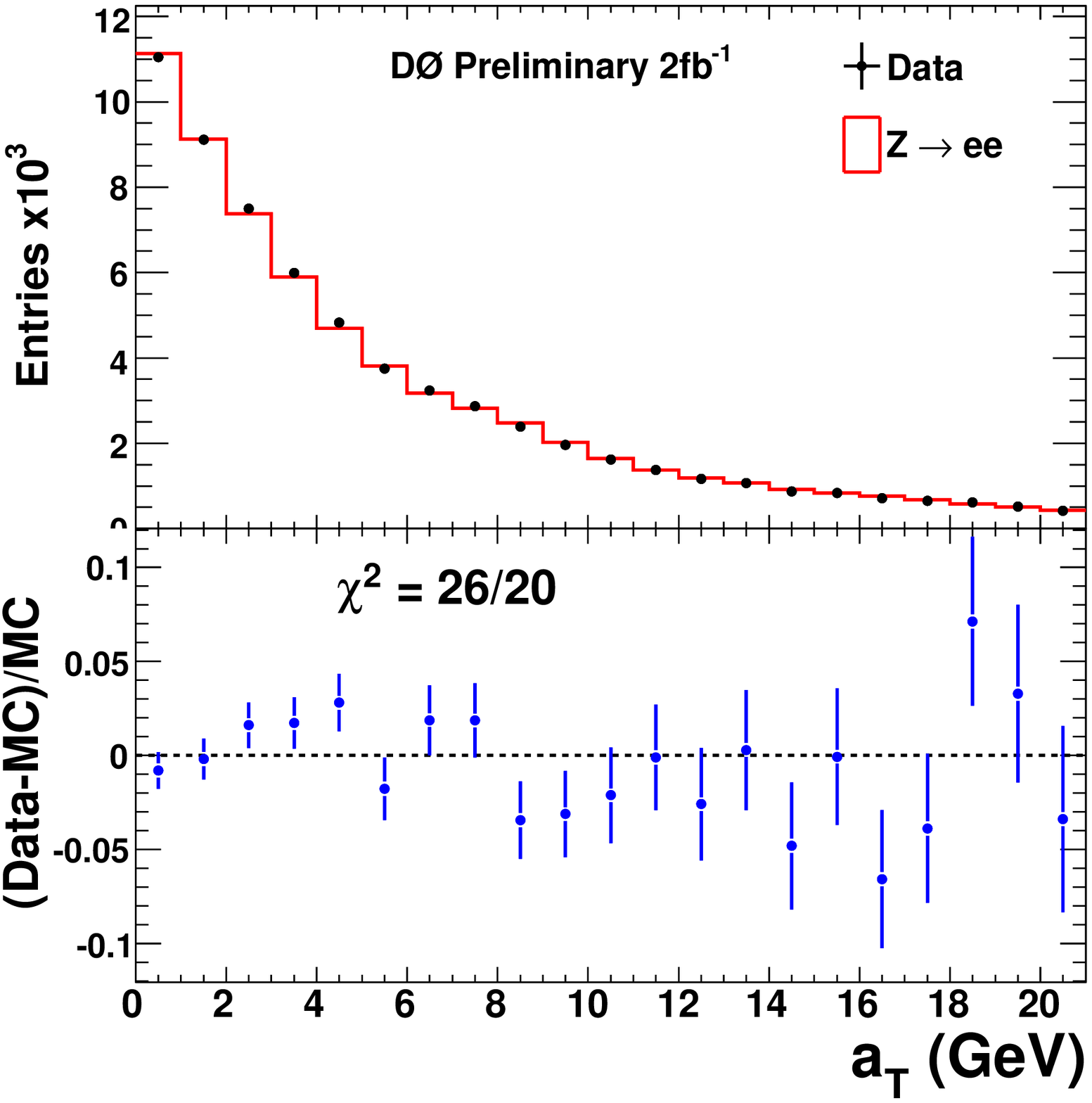}
\includegraphics[width=85mm]{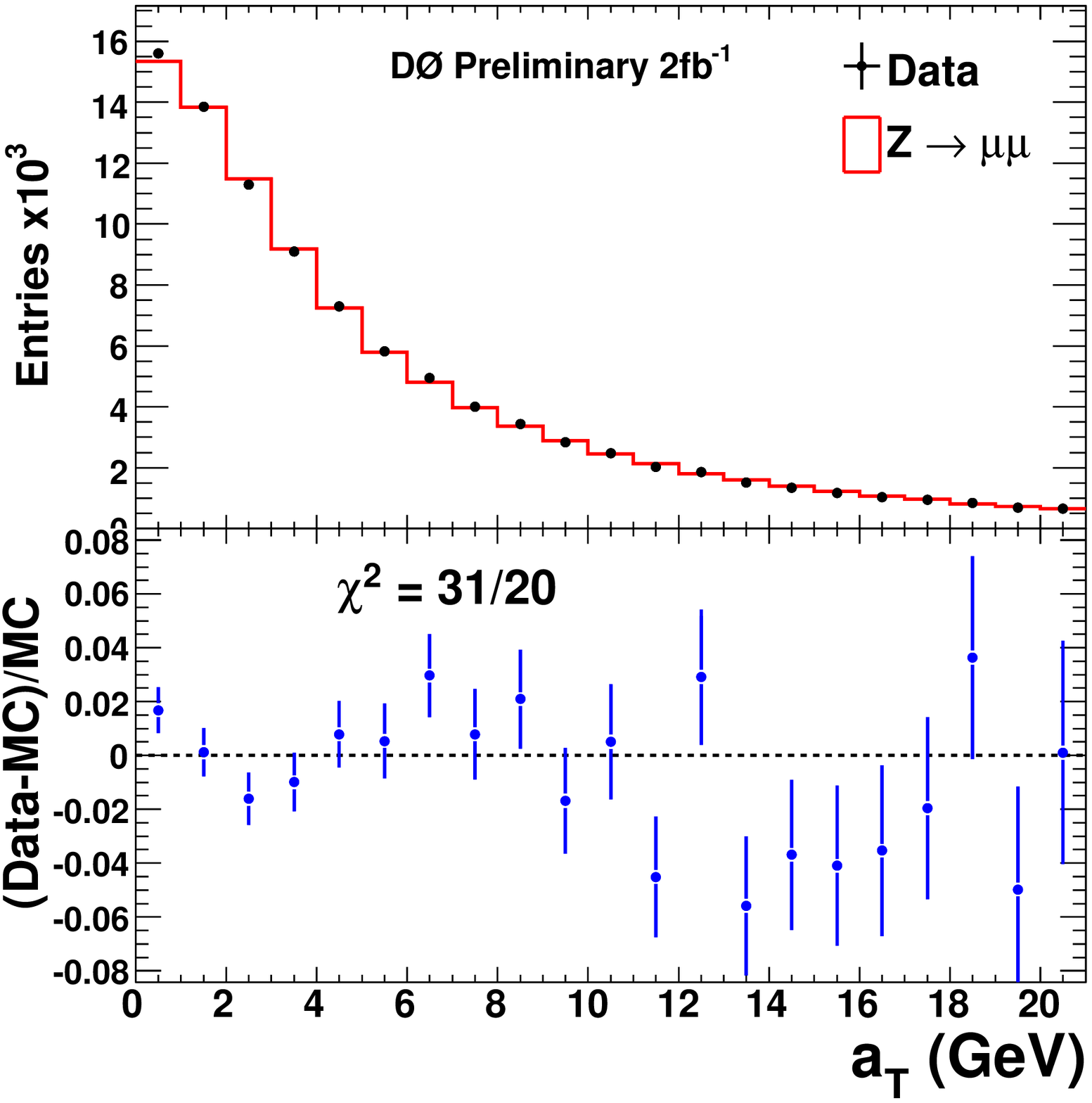}
\caption{The measured dimuon $p_T$ projected along the axis bisecting the muon pair ($a_T$), 
for 2 fb$^{-1}$ of dielectron and dimuon D\O\ data, and for {\sc resbos} simulation using 
CTEQ6.6 and $g_2 = 0.63$. }
\label{fig:d0zat}
\end{figure}

D\O\ has performed a dedicated measurement of $g_2$ for use in its $m_W$ measurement.
To maximize sensitivity to $g_2$, D\O\ projects the $Z$ boson $p_T$ along the axis 
bisecting the charged leptons (Fig.~\ref{fig:d0zat}).  Fitting this distribution in the  
electron and muon decay channels gives $g_2 = 0.63 \pm 0.02$ using CTEQ6.6 PDFs.  D\O\ 
has studied the PDF uncertainty on $g_2$, finding $\delta g_2 {\rm (PDF)} = 0.04$.   

The crucial step for the $m_W$ measurement is translating the $g_2$ value obtained by 
fitting the $Z$ boson $p_T$ into the appropriate value for $p_T^W$.  The {\sc resbos} 
parametrization provides this translation, but the uncertainty due to higher resummation 
orders has not been determined.  In addition, variations in $\alpha_s$ affect the high 
end of the $p_T$ spectrum and could cause a small uncertainty on $m_W$.  At the low end 
of the spectrum, small modifications could arise from QED initial-state radiation, which 
should be investigated.  Other issues relevant to the $m_W$ measurement are including the 
full correlations with PDFs when determining the uncertainty, and including the diffractive 
production component that is not modelled by {\sc resbos}.

Starting with the D\O\ Run 1B $m_W$ measurement \cite{d01b}, the Tevatron experiments have 
quoted measurements of $m_W$ based on fits to the charged-lepton and neutrino $p_T$ distributions.
These fits are more sensitive to the modelling $p_T^W$ than the traditional $m_T$ 
($=\sqrt{2p_T^l p_T^{\nu}(1 - \cos\Delta\phi)}$) fit, providing an important test of the 
$p_T^W$ model.  

%%%%%%%%%%%%%%%%%%%%%%%%%%%%%%%%%%
\subsection{QED Radiation}

Final-state photon radiation (FSR) off a charged lepton from the $W$ boson decay reduces 
the charged lepton momentum, and thus the inferred boson mass.  Modelling this effect 
results in an ${\cal {O}}(150$ MeV) correction to the measured $m_W$ \cite{cdfwmasswidth}.  
D\O\ models FSR with {\sc photos} \cite{photos}, a resummed calculation focused exclusively 
on FSR, and compares with {\sc wgrad} \cite{wgrad}, a next-to-leading order (NLO) calculation.  
CDF uses a histogram of photons extracted from {\sc wgrad} to apply FSR at the end of event 
generation.  

To determine the uncertainty on $m_W$ due to the FSR model, D\O\ takes the difference 
between fits using {\sc photos} and {\sc wgrad}.  This is almost certainly an overestimate, 
since {\sc photos} includes higher-order terms (through resummation) that {\sc wgrad} does 
not.  The uncertainty is nonetheless small due to the electron energy calibration using the 
$Z$ boson mass, which largely corrects for mismodelling of photon radiation.

CDF models higher-order QED radiation by scaling the photon energy by 10\%, taking half 
the scaling correction as an uncertainty.  Other uncertainties due to the infrared cutoff 
in {\sc wgrad} and a comparison of full ${\cal{O}}(\alpha)$ and FSR-only {\sc wgrad} are 
also quoted.  CDF is undertaking a thorough investigation of higher-order QED effects using 
the {\sc horace} generator \cite{horace}.

{\sc horace} calculates the leading logarithm QED corrections, and reweights them to 
model the full $\alpha^n$ calculation.  The procedure assumes that the reweighting needed 
to model ${\cal{O}}(\alpha)$ is the same for all orders of $\alpha$.  A CDF study has found 
that the reweighting produces a $4.5 \pm 1.4$ MeV shift in the $m_W$ fit, when compared to 
the leading log calculation.  A comparison of the reweighted logs with the 
${\cal{O}}(\alpha)$ calculation shows an $\approx 10$ (20) MeV shift for electrons (muons). 
The mass shift is lower in the reweighted log simulation, since the higher orders 
suppress soft QED radiation.  Variations in the truncation of the perturbative series in 
{\sc horace} have less than a 1 MeV effect on $m_W$.  

The {\sc horace} generator improves our understanding of QED radiation, though there is 
no clear recipe for determining the residual uncertainty on $m_W$.  A full 
${\cal{O}}(\alpha^2)$ calculation would be useful in order to validate the {\sc horace} 
reweighting procedure, but this requires significant effort.  Such a calculation could 
address the question of whether there are uncertainties due to additional diagrams not 
accounted for in the {\sc horace} reweighting scheme (e.g., final-state radiation of 
electron-positron pairs).  Alternative generators could also be useful; for example, the 
{\sc winhac} \cite{winhac} generator incorporates higher orders through exponentiation 
rather than showering and could provide a cross-check, but not a measure of uncertainty.  

Another issue is the potential correlation between initial-state QCD and final-state 
QED radiation.  Currently, CDF and D\O\ factorize the two, generating QED FSR after 
{\sc resbos}, or boosting bosons produced by {\sc horace}.  Recently, unified 
generators have become available: the {\sc horace} authors have added {\sc mc@nlo} \cite{mcnlo} 
for QCD ISR, and the {\sc resbos} authors have added QED FSR in the generator {\sc resbosa}.  
However, these are still factorized approaches and do not include interference between 
QCD and QED radiation.

Uncertainties on QED FSR can be mitigated by calibrating the lepton momentum using the 
$Z$ boson mass in $Z \rightarrow ll$ events.  D\O\ uses this technique for its calibration, 
though CDF does not because doing so would inflate the overall uncertainty due to the 
relatively small $Z \rightarrow ll$ statistics. 

%%%%%%%%%%%%%%%%%%%%%%%%%%%%%%%%%%
\subsection{Boson Decay}

The left-handed coupling of the $W$ boson to the quarks and leptons produces a decay 
angular distribution proportional to $(1 + \cos\theta)^2$ for production by valence 
quarks at leading order, where $\theta$ is the angle between the (anti)quark and 
(anti)lepton momenta.  Higher-order QCD corrections modify the angular distributions, 
and have been calculated at NLO and implemented in {\sc resbos}.  A comparison of 
{\sc resbos} to the dedicated NLO generator {\sc dyrad} \cite{dyrad} shows consistency in 
the region of high $W$ boson $p_T$ ($p_T^W > 15$ GeV).  At lower $p_T$ the distributions 
are more accurately described by a resummation procedure, which for {\sc resbos} involves 
an averaging over helicities.  Resummation calculations separated by helicity are in 
progress, but until they are complete there is some ambiguity of the uncertainty on 
$m_W$ from the {\sc resbos} decay model.  

%%%%%%%%%%%%%%%%%%%%%%%%%%%%%%%%%%
\section{Tevatron $m_W$ Measurements}

The CDF and D\O\ experiments use independent procedures to calibrate the detector 
response to charged leptons and to hadrons from the underlying event.  CDF utilizes its 
precision tracker to measure $m_W$ in both $W\rightarrow \mu\nu$ and $W\rightarrow e\nu$ 
decays, while the D\O\ measurement relies on its hermetic calorimeter to focus on the 
electron decay channel. 

%%%%%%%%%%%%%%%%%%%%%%%%%%%%%%%%%%
\subsection{Charged Lepton Calibration} 
The CDF lepton momentum calibration begins with the tracker.  Charged-track momentum 
is calibrated using $J/\psi\rightarrow\mu\mu$, $\Upsilon\rightarrow\mu\mu$, and 
$Z \rightarrow \mu\mu$ events.  Fits to the invariant mass of muon pairs in these 
samples set the momentum scale, and are sensitive to modelling of the ionization 
energy loss.  CDF models the energy loss using the mean from the Bethe-Bloch equation \cite{pdg} 
for each traversed layer of material.  In the $J/\psi$ sample, which contains more than 
600,000 events, the calibration uncertainty is dominated by the energy loss model.  
Modelling the energy loss as a Landau distribution could improve the quality of the 
calibration fit, resulting in a smaller overall uncertainty.  However, some care is 
required to preserve the Bethe-Bloch mean when using the Landau distribution.

\begin{figure}[h]
\centering
\includegraphics[width=85mm]{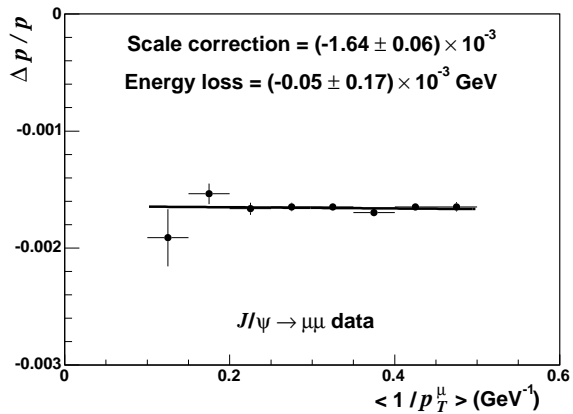}
\caption{The momentum scale required in the simulation to obtain the world-average 
$J/\psi$ mass.  The scale is plotted as a function of mean inverse $p_T$ of muon pairs 
from $J/\psi$ decays, and fit to a line where the slope equals the residual energy loss 
and the intercept equals the momentum scale. }
\label{fig:jpsi}
\end{figure}

CDF calibrates the average muon energy loss by fitting the momentum scale as a function 
of mean inverse transverse momentum of muons from $J/\psi$ decays (Fig.~\ref{fig:jpsi}).  
The width of the observed peak is also sensitive to the intrinsic detector resolution 
and multiple scattering in the detector.  The multiple scattering model includes 
non-Gaussian tails based on data from low-energy muons impinging on a fixed target \cite{muscat}.

\begin{figure}[h]
\centering
\includegraphics[width=90mm]{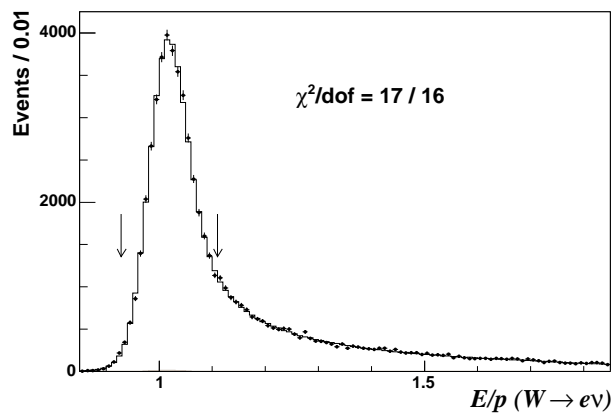}
\caption{The ratio of calorimeter energy to track momentum for electrons from $W$ boson 
decays.  The region between the arrows is used to fit for the calorimeter energy scale. }
\label{fig:eop}
\end{figure}

The CDF electron momentum calibration transfers the track calibration to the calorimeter 
using electrons from $W$ boson decays.  CDF fits the ratio of calorimeter energy to 
track momentum ($E/p$) using the peak region (Fig.~\ref{fig:eop}).  The position of the 
peak is sensitive to the radiation of low-momentum photons in the tracker.  The rate 
of this radiation is in turn sensitive to the amount of tracker material, which is tuned 
using the high end of the $E/p$ ($1.19-1.85$) distribution.  This tuning empirically 
corrects the rate of high-momentum radiation, and relies on the theoretical radiation 
spectrum to model the region near the peak.  For very low momentum radiation 
($\lesssim 50$ MeV), the model includes quantum-mechanical interference effects that 
suppress the radiation.  

CDF tests its lepton momentum calibration by fitting for the $Z$ boson mass using 
$Z\rightarrow ll$ events.  The consistency of the fit $m_Z$ with the LEP measurements 
\cite{ewkmeasurements} in both the electron and muon decay channels provides a stringent 
test of the detector response and modelling of radiative corrections from first principles.
CDF adds the $m_Z$ fit to its momentum calibration, though the relatively low $Z$ boson 
statistics results in a negligible contribution to the muon calibration, and a 30\% contribution 
to the electron calibration.  Since the muon calibration uncertainties are dominantly 
systematic, it is expected that future measurements will rely more on the $Z$ boson mass fit.  

\begin{figure}[h]
\centering
\includegraphics[width=75mm]{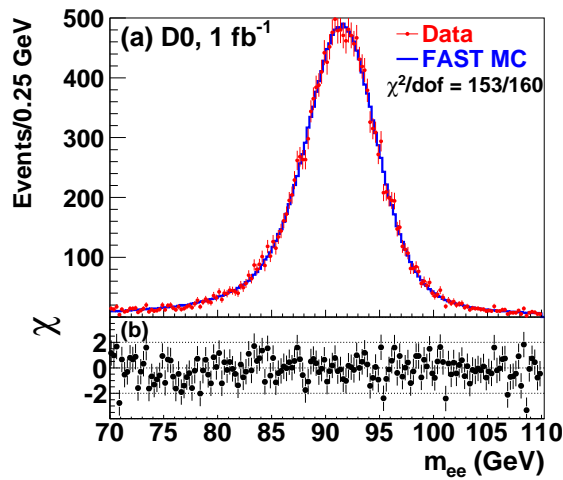}
\caption{(a) The dielectron mass distribution for 1 fb$^{-1}$ of calibrated D\O\ data and 
simulation. (b) The difference between data and simulation, divided by the expected statistical 
uncertainty on the data. }
\label{fig:d0zmass}
\end{figure}

The D\O\ electron momentum calibration is based solely on fits to the $Z$ boson mass, as 
a function of detector region.  The calibration determines both the energy scale and an 
offset.  The offset corrects for any inaccuracies in the modelling of detector noise 
and underlying event in the electron energy measurement.  After calibration, the $Z$ boson 
mass distribution is well described by the simulation (Fig.~\ref{fig:d0zmass}).  

The use of $Z \rightarrow ll$ events for calibration cancels a number of systematic 
uncertainties when applied to the $W \rightarrow l\nu$ mass sample.  However, there is 
no independent test of the calibration, making the measurement less robust and ultimately 
increasing the overall uncertainty.  In addition, extra care must be taken to understand and 
account for uncertainties that do not cancel when the calibration is applied to the $W$ boson 
sample.  For example, the electron $p_T$ measurement relies on a measurement of the track angle 
with respect to the beam line.  A global scale of this angle that brings the track closer 
to the beam line can bias the $m_Z$ fit up or down, depending on the topology.  However, the 
$m_W$ fit is always biased to lower values.

%%%%%%%%%%%%%%%%%%%%%%%%%%%%%%%%%%
\subsection{Neutrino Calibration} 
\begin{figure}[h]
\centering
\includegraphics[width=75mm]{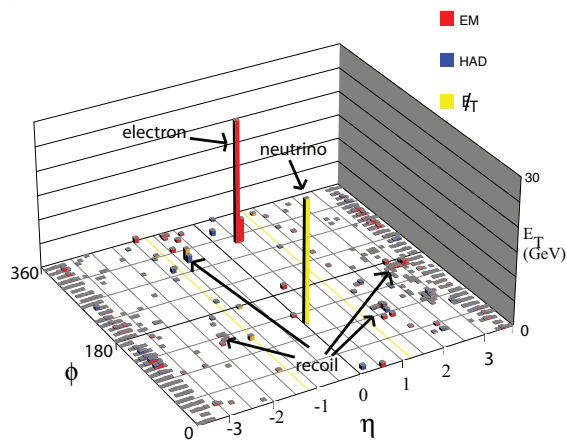}
\caption{A $W\rightarrow e\nu$ candidate event collected with the D\O\ detector.  The 
electron and representative components of the recoil are indicated, as well as the 
inferred neutrino position and transverse momentum.  }
\label{fig:recoil}
\end{figure}

Since the neutrino momentum is inferred from the measured momentum imbalance of the 
event, the neutrino calibration is effectively a calibration of all the particles in 
the event.  Excluding the charged lepton from the $W$ boson decay, these particles are 
known as the recoil.  The recoil momentum is the vector sum of diffuse contributions 
that are not as well measured as the charged lepton (Fig.~\ref{fig:recoil}).  The 
detector response to these particles uses an empirical model with parameters determined 
from $Z \rightarrow ll$ events.

\begin{figure}[ht]
\centering
\includegraphics[width=90mm]{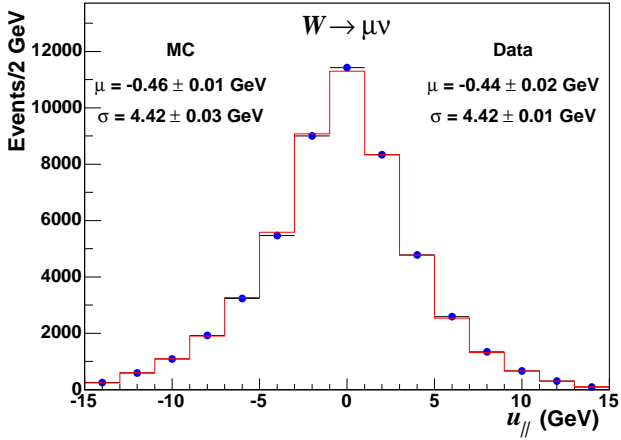}
\caption{The projection of the recoil vector along the direction of the muon in 
$W\rightarrow \mu\nu$ events collected by the CDF detector.  A bias in the mean of this 
distribution translates into an equivalent bias on the $W$ boson mass measured using the 
$m_T$ distribution. }
\label{fig:cdfwupar}
\end{figure}

The CDF recoil calibration defines a physics-motivated model for the scale and resolution 
of the recoil.  The scale is a logarithmic function of boson $p_T$.  As $p_T$ increases, 
the particles in the recoil have higher $p_T$ and are contained in a smaller jet cone.  The 
resolution improves with increasing $p_T$, with the expected dependence of a sampling 
calorimeter.  The resolution due to underlying event uses the same dependence, with 
parameters determined from data collected with an unbiased trigger.  With a few parameters 
CDF has demonstrated quantitative agreement between data and simulation for the important 
recoil distributions in the $W$ boson samples.  For example, the projection of the recoil 
vector along the direction of the muon in $W\rightarrow \mu\nu$ events is shown 
Fig.~\ref{fig:cdfwupar}.  

\begin{figure}[ht]
\centering
\includegraphics[width=85mm]{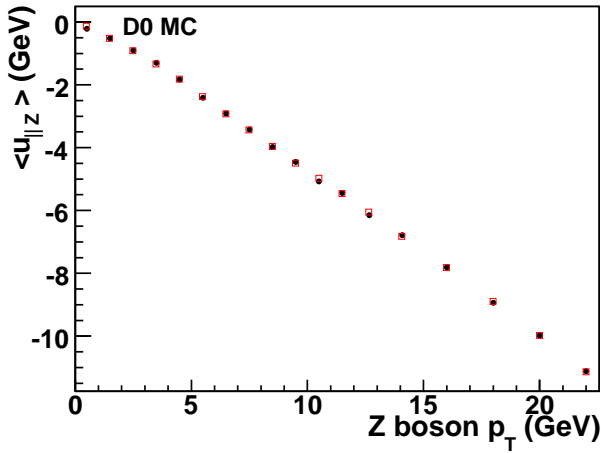}
\caption{The projection of the recoil vector along the direction of the electrons in 
$Z\rightarrow ee$ events collected by the D\O\ detector. }
\label{fig:d0zupar}
\end{figure}

The D\O\ recoil calibration uses a detector-response library as a function of true recoil 
momentum, derived using $Z\rightarrow ll$ events and an unbiased trigger to model the effects 
of the underlying event \cite{d0recoil}.  The use of a library removes any assumptions on the 
form of the response functions, though the properties of these functions must be checked to 
ensure that they are realistic.  The D\O\ model accurately describes the recoil distributions 
of the calibration sample; as an example, the projection of the observed recoil along the 
direction of the electrons in $Z\rightarrow ee$ events is shown in Fig.~\ref{fig:d0zupar}.  
The quality of the model for the $W$ boson recoil distributions has yet to be shown.

%%%%%%%%%%%%%%%%%%%%%%%%%%%%%%%%%%
\subsection{$W$ Boson Mass Fits} 
\begin{figure}[t]
\centering
\includegraphics[width=90mm]{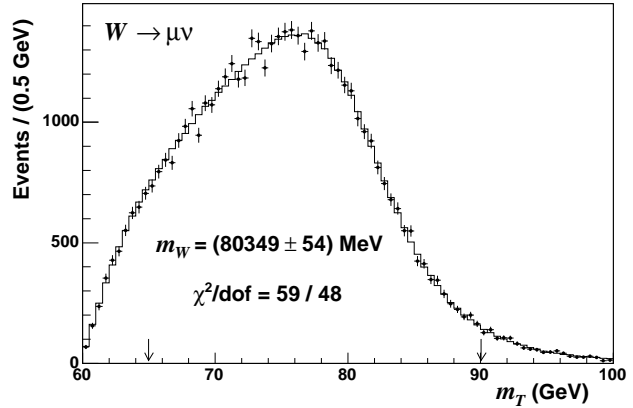}
\includegraphics[width=90mm]{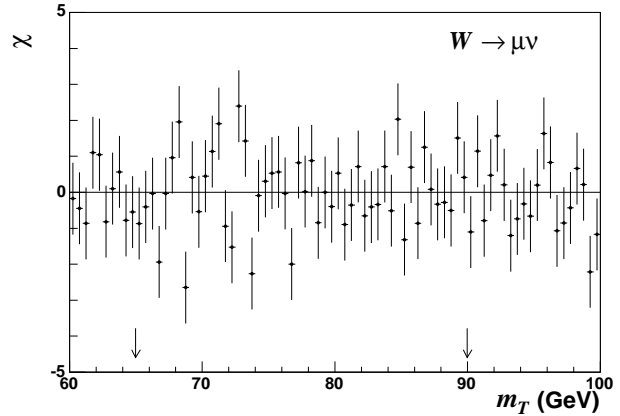}
\caption{Top: The fit to the $m_T$ spectrum from the $W\rightarrow\mu\nu$ decays measured 
with the CDF detector.  Bottom: The difference between data and simulation, divided by the 
statistical uncertainty on the expectation. }
\label{fig:cdfmassfits}
\end{figure}

\begin{figure}[t]
\centering
\includegraphics[width=80mm]{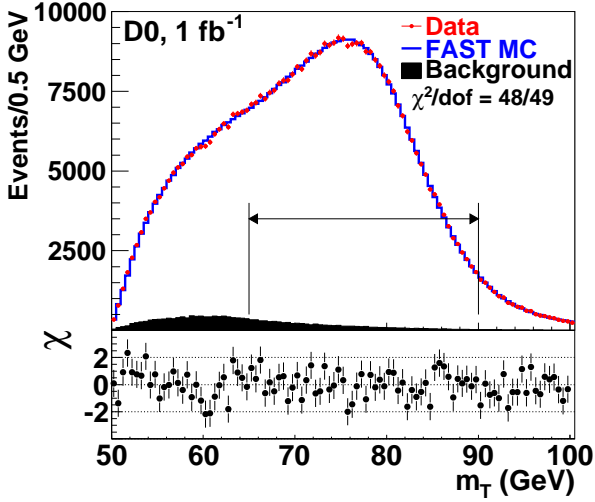}
\caption{The projection of the recoil vector along the direction of the electrons in 
$Z\rightarrow ee$ events collected by the D\O\ detector. }
\label{fig:d0massfits}
\end{figure}

Fits for $m_W$ are performed using the charged-lepton and neutrino $p_T$ spectra, and the 
reconstructed $m_T$ distribution.  The latter provides the most precise measurement of $m_W$, 
with the combination of the fits improving the total precision by a few percent.  The results 
of the $m_T$ fits in the muon channel at CDF and the electron channel at D\O\ are shown in 
Figs.~\ref{fig:cdfmassfits} and \ref{fig:d0massfits}, respectively.  

The CDF and D\O\ $m_W$ measurements are:
\begin{eqnarray}
m_W & = & 80.413 \pm 0.034{\rm (stat)} \pm 0.034{\rm (sys)} {\rm ~GeV} {\rm ~(CDF),} \nonumber \\
m_W & = & 80.401 \pm 0.021{\rm (stat)} \pm 0.038{\rm (sys)} {\rm ~GeV} {\rm ~(D\O\ \! \!).} \nonumber 
\end{eqnarray}

\noindent
These are the two most precise measurements from individual experiments.  The systematic 
uncertainties on the measurements are shown in Tables~\ref{tbl:d0uncertainty} and 
\ref{tbl:cdfuncertainty}.  In both cases the dominant uncertainty is on the lepton 
momentum scale calibration, which is performed in situ.  Thus, this uncertainty is 
expected to reduce with increased statistics.

\begin{table}[htbp]
\begin{center}
  \begin{tabular}{ lccc}
          & \multicolumn{3}{c}{$\delta m_W$~(MeV)} \\
   Source                          &$m_T$ & $p_T^e$ &  $p_T^{\nu}$\\
  \hline \hline
  Electron energy calibration       & 34 &  34 & 34 \\
  Electron resolution model         &  2 &   2 &  3 \\
  Electron shower modeling           &  4 &   6 &  7 \\
  Electron energy loss model        &  4 &   4 &  4 \\
  Hadronic recoil model             &  6 &  12 & 20 \\
  Electron efficiencies             &  5 &   6 &  5 \\
  Backgrounds                       &  2 &   5 &  4 \\ \hline
  Experimental Subtotal             & 35 &  37 & 41 \\ \hline

  PDF                          &  10 &  11 & 11 \\
  QED                          &  7 &   7 &  9 \\
  Boson $p_T$                  &  2 &   5 &  2 \\ \hline
  Production Subtotal          & 12 &  14 & 14 \\ \hline

  Total                        &  37 & 40 & 43 \\
  \end{tabular}
  \caption{Systematic uncertainties on the D\O\ $m_W$ measurement \cite{d0wmasswidth}.}
\label{tbl:d0uncertainty}
\end{center}
\end{table}

\begin{table}[htbp]
\begin{center}
\begin{tabular}{lc}
\hline
\hline
Source                   & Uncertainty (MeV) \\
\hline
Lepton Scale             & 23.1 \\
Lepton Resolution        & 4.4 \\
Lepton Efficiency        & 1.7 \\
Lepton Tower Removal     & 6.3 \\
Recoil Energy Scale      & 8.3 \\
Recoil Energy Resolution & 9.6 \\
Backgrounds              & 6.4 \\
PDFs                     & 12.6 \\
$W$ Boson $p_T$          & 3.9 \\
Photon Radiation         & 11.6 \\
\hline
\hline
\end{tabular}
\caption{Systematic uncertainties on the combination of the six fits in the electron and 
muon channels for the CDF $m_W$ measurment \cite{cdfwmasswidth}. } 
\label{tbl:cdfuncertainty}
\end{center}
\end{table}

Combining all Tevatron measurements gives \cite{combination}:
\begin{equation}
m_W = 80.420 \pm 0.031 {\rm ~GeV} {\rm ~(Tevatron),}\nonumber \\
\end{equation}

\noindent
which is more precise than the combined LEP measurement of $m_W = 80.376 \pm 0.033$ GeV.  
The current world-average value of $m_W$ is \cite{combination}
\begin{equation}
m_W = 80.399 \pm 0.023 {\rm ~GeV} {\rm ~(World~average).}\nonumber \\
\end{equation}

%%%%%%%%%%%%%%%%%%%%%%%%%%%%%%%%%%
\section{Global Electroweak Fits}
\begin{figure}[t]
\centering
\includegraphics[width=80mm]{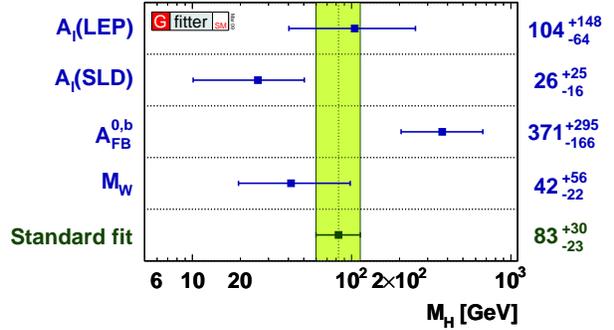}
\caption{The value of $m_H$ preferred by measurements of the forward-backward lepton asymmetry in 
$e^+ e^-$ collisions at LEP and SLD, the forward-backward $b-$quark asymmetry at SLD, and the 
$m_W$ measurement.  All variables sensitive to $m_H$ are removed from the fit except for the one 
quoted \cite{gfitter}. }
\label{fig:gfitter}
\end{figure}

Several groups have updated their fits to the global electroweak data using the latest $m_W$ 
measurements.  The Gfitter group has obtained a best-fit Higgs boson mass $m_H = 83^{+30}_{-23}$ 
GeV \cite{gfitter}, more than $1\sigma$ below the LEP direct exclusion $m_H > 114$ GeV \cite{lephiggs}.  
There is thus tension between the electroweak fits and $m_H$, and this tension increases if one only 
considers the predictions from $m_W$ alone.  The Gfitter group has fit for $m_H$ using only one 
sensitive variable at a time, and obtains $m_H = 42^{+56}_{-22}$ GeV when using only $m_W$ 
(Fig.~\ref{fig:gfitter}).  In fact, all measurements prefer a low-mass Higgs, except the 
forward-backward asymmetry measurement in polarized electron-positron collisions.  The Gfitter 
group has determined the probability of such a deviant measurement to be 1.4\%, if due to 
measurement uncertainties alone.

\begin{figure}[t]
\centering
\includegraphics[width=80mm]{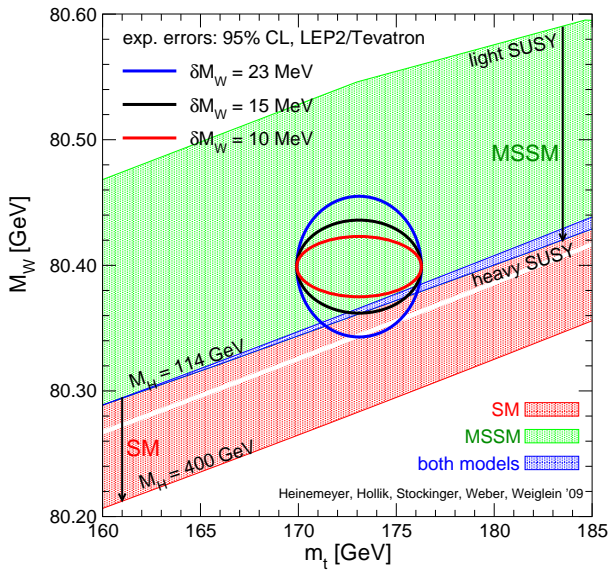}
\caption{The 95\% ellipse for the measured values of $m_W$ and $m_t$ (blue circle), and smaller 
ellipses for future $m_W$ uncertainties of 15 MeV (black circle) and 10 MeV (red circle) 
\cite{heinemeyer}.  An uncertainty of 10 MeV would exclude values of the SM Higgs above the 
current direct lower limit of 114 GeV. }
\label{fig:mssm}
\end{figure}

Given the tension between the electroweak fits and the direct limit on $m_H$, it is natural 
to consider possible new-physics contributions to $m_W$.  One possibility is the presence of 
supersymmetric-particle loops in the $W$ boson propagator.  Such loops in the minimal supersymmetric 
standard model increase the $W$-boson mass and reduce this tension.  Figure~\ref{fig:mssm} shows 
the range of top-quark and $W$-boson masses preferred by the MSSM and by the SM \cite{heinemeyer}.  
However, there are other constraints on supersymmetry that create a different set of tensions.  

%%%%%%%%%%%%%%%%%%%%%%%%%%%%%%%%%%
\section{Future Measurements}
In addition to the expected improvement in $m_W$ precision from analysis of the complete Tevatron 
datasets, even greater precision is predicted by the ATLAS experiment at the LHC.  With 10 fb$^{-1}$ 
of $\sqrt{s} = 14$ TeV data (one year of running at design luminosity), the ATLAS experiment expects 
to have a precision of 7 MeV on its measurement of $m_W$ \cite{atlas}.  However, there are significant 
challenges to achieving this goal.  For example, the ATLAS projection is based on the charged-lepton 
$p_T$ fit for a single decay channel and assumes the recoil uncertainty is negligible.  However, both 
the CDF and D\O\ measurements find a larger recoil uncertainty on this mass fit than on the fit to the 
$m_T$ distribution, due to the tight cut on the recoil momentum in the event selection.  Thus, one 
would expect a non-negligible recoil uncertainty for the ATLAS measurement.  In addition, ATLAS expects 
the $p_T^W$ uncertainty to dominate the production model uncertainty, with a negligible PDF uncertainty.
Given the large uncertainties on the PDFs at the momentum fraction relevant for $W$ boson production 
at the LHC, this projection appears optimistic.  

Even though there will be significant challenges to overcome for measuring $m_W$ at ATLAS and CMS, 
there will be ${\cal{O}}(10^8)~W$- and ${\cal{O}}(10^7)~Z$-boson events to calibrate the detector 
response to leptons and recoil, and to measure the $Z$ boson rapidity and $p_T$ distributions to 
constrain the PDFs and $p_T^W$.  In addition, there is a $W$ boson charge asymmetry at the LHC which 
is similar to that of the Tevatron, providing further possible PDF constraints.  Given these large 
statistics and calibration tools, it is realistic to expect a measurement with precision better than 
10 MeV from the LHC experiments.

Finally, there is potential for a precision measurement of the weak mixing angle from the 
forward-backward asymmetry of leptons in Drell-Yan production at the Tevatron.  The distribution 
of the angle $\theta$ between the negative-lepton and proton momenta has the form \cite{cdfafb}:
\begin{equation}
d\sigma / d\cos\theta \propto 3/8(1 + \cos^2\theta) + A_{FB}\cos\theta,
\end{equation}

\noindent
where $A_{FB}$ is the asymmetry between negative leptons produced in the forward ($\cos\theta > 0$) 
and backward ($\cos\theta < 0$) directions, and is a function of the vector and axial couplings 
of the fermions to the $Z$ and $\gamma$ bosons.  Since the vector coupling is equal to 
$I^3_L - 2e\sin^2\theta_W$, where $I^3_L$ is the weak charge, the measurement provides sensitivity 
to the weak mixing angle.

CDF and D\O\ have performed measurements of $A_{FB}$ in the electron decay channel with 72 pb$^{-1}$ 
\cite{cdfafb} and 1.1 fb$^{-1}$ \cite{d0afb} of data, respectively.  Assuming statistical scaling of 
the uncertainties on the D\O\ measurement, 
\begin{equation}
\sin^2\theta_W = 0.2326 \pm 0.0018{\rm (stat)} \pm 0.006{\rm (sys)}, 
\end{equation}

\noindent
the combined Tevatron precision using electron and muon channels in ${\cal{O}}(10)$ fb$^{-1}$ of data 
could approach 0.0003, which would contribute to the world-average value of 
$\sin^2\theta_W = 0.23149 \pm 0.00013$.  

%%%%%%%%%%%%%%%%%%%%%%%%%%%%%%%%%%
\section{Conclusions}
The model of electroweak unification has been tested to high precision, and is now used to constrain 
the existence and properties of unobserved particles coupling to the $W$ and $Z$ bosons.  Further 
progress relies on improving the measurement of the $W$ boson mass, whose uncertainty is the limiting 
factor in these constraints.  Measurements at the Tevatron have recently reduced this uncertainty 
significantly, with further reduction expected from analysis of the existing data.  In order to achieve 
the stated goal of a measurement more precise than the current world average, the Tevatron experiments 
require improvements in the model of $W$ and $Z$ boson production.  Ongoing theoretical work and recent 
production measurements should produce the needed improvements.  Ultimately, the LHC should produce a 
measurement of $m_W$ with better than 10 MeV precision.  At that point, other uncertainties on parameters 
in the electroweak fits will need to be revisited, for example the uncertainty on the electromagnetic 
coupling $\alpha$, evaluated at the $Z$ boson mass.

Another possibility for improving electroweak constraints is the determination of $\sin^2\theta_W$ 
through the measurement of the forward-backward asymmetry of Drell-Yan at the Tevatron.  However, 
much work is required to demonstrate the scaling of uncertainties with an order of magnitude more 
data, and to achieve sensitivity in the muon decay channel.

Overall, there is significant ongoing progress in precision electroweak measurements at hadron 
colliders.  The constraints on the Higgs boson mass are quickly tightening, and if there is no 
SM Higgs there is a reasonable possibility that it will be first excluded by the $W$ boson mass 
measurement.

%%%%%%%%%%%%%%%%%%%%%%%%%%%%%%%%%%
\begin{acknowledgments}
I would like to thank the Electroweak session conveners, Ashutosh Kotwal, Uli Baur, and Tim Bolton, 
for inviting me to give this review, and for providing useful comments on these proceedings.
\end{acknowledgments}

\bigskip % extra skip inserted

\end{document}